\documentclass{JHEP3}
\usepackage{amsmath,amssymb,amsfonts,mathrsfs}
\usepackage{url}
\usepackage{graphicx}
\usepackage[all]{xy} 
\usepackage[vcentermath]{youngtab}
\usepackage{young}

\newtheorem{thm}{Theorem}[subsection]

\newtheorem{prop}[thm]{Proposition}
\newtheorem{rem}[thm]{Remark}
\newtheorem{defn}[thm]{Definition}

\newcommand{\C}{\mathbb{C}}

\newcommand{\Z}{\mathbb{Z}}

\newcommand{\calI}{\mathcal{I}}
\newcommand{\vecR}{\vec{R}}

\title{Notes On U(1) Instanton Counting On $A_{l-1}$ ALE Spaces}

\author{Haitao Liu\\
  Department of Mathematics and Statistics,\\
  University of New Brunswick, Fredericton, Canada, E3B 5A3\\
  and\\
  Department of Applied Mathematics, \\
  Hebei University of Technology, Tianjin, P.R.China, 300130\\
  Email: \email{haitao.liu@unb.ca}\\}

\abstract{In this note, we investigate the detailed relationship between the orbifold partition counting and the (l-quotient, l-core) pair counting. We show that the orbifold partition counting is exactly the same as the (l-quotient, l-core) pair counting. } 


\preprint{}

\begin{document}
\section{Introduction}
In recent years, people have made much progress on connecting the four-dimensional supersymmetric gauge theories with the two dimensional conformal theories, e.g. \cite{Nekrasov:2009rc, Alday2010, Nekrasov:2010ka}.  One of the first examples was found by Nakajima \cite{Nakajima1994}. Further, in \cite{Vafa1994}, Witten and Vafa discussed this in four-dimensional supersymmetric gauge theory based on the ALE space by using the twisting method. Nakajima's analysis shows that the partition functions of N=4 four-dimensional U(N) gauge theories on the $A_{l-1}$ ALE space is related to the affine character of $\widehat{su}(l)_N$. Since the boundary of $A_{l-1}$ ALE space is the lens space $S^3/\Z_l$, so the U(N) gauge theory can approach some non-trivial flat connection at infinity \cite{Griguolo:2006kp, Dijkgraaf2008a}. Such flat connections are labeled by the N-dimensional representation of $\Z_k$ 
\begin{equation*}
\lambda\in Hom(\Z, U(N)),
\end{equation*}
which can be decomposed into irreducible representations $R_a$ of $\Z_l$ as 
\begin{equation*}
\lambda=\sum_{a}R_aN_a,
\end{equation*}
where $N_a$ are the integers satisfying 
\begin{equation*}
\sum_a N_a\text{dim}R_a=N.
\end{equation*}
Thus once we choose boundary condition $\lambda$ at infinity we get a vector-valued partition function whose components are of the form
\begin{equation}
\mathcal{Z}_\lambda(v, \tau)=\chi_{\hat{\lambda}}^{\widehat{su}(l)_N}(v, q).
\end{equation}
Witten and Vafa showed that the partition function of N=4 U(N) super-Yang-Mills theory on the ALE spaces is related to the generating function of the Euler number of the instanton moduli spaces \cite{Vafa1994}. Later, in \cite{Fucito2004}, the authors employed equivariant cohomology techniques to calculate the partition function explicitly.  Recently Dijkgraaf and Su{\l}kowski introduce the orbifold partitions and show how to get the affine character of $\widehat{su}(l)_1$ for the U(1) gauge theory \cite{Dijkgraaf2008a}. 

In this note, we investigate the detailed relationship between the orbifold partition counting and the (l-quotient, l-core) pair counting. In section 2, we review the definition of orbifold partitions introduced in \cite{Dijkgraaf2008a}. In section 3, we review the basic structures of U(1) instantons on $A_{l-1}$ ALE spaces. In section 4, we investigate the detailed relation between the orbifold partition counting and the (l-quotient, l-core) pair counting and  show that the orbifold partition counting is exactly the same as the (l-quotient, l-core) pair counting.

\section{Orbifold partitions}
In this section we briefly review the first type of orbifold partition counting in \cite{Dijkgraaf2008a}. 

We know that an ideal of functions $\calI=\{f(x,y)\}\subset \C[x,y]$ generated by a set of monomials $x^iy^j$ for $i, j\geq 0$ corresponds to an ordinary two-dimensional partition $\lambda$ in such a way that a box $(m,n)\in\lambda$ iff $x^my^n\notin\calI$. For example, the following Young diagram shows the $\calI$ which is generated by $y^4, xy^2, x^2y, x^2$. 
\begin{equation*}
\begin{Young}
$y^3$\cr
$y^2$\cr
$y$&$xy$\cr
$1$&$x$\cr
\end{Young},
\label{eq01}
\end{equation*}

For the $A_{l-1}$ ALE space we will consider ideals of functions having definite transformation properties under the action
\begin{equation}
(x, y)\rightarrow (\omega x, \bar{\omega}y),  \label{eq1}
\end{equation}
where $\omega=e^{2\pi i/l}$. All the monomials with the same transformation property form a periodic sub-lattice of $\Z^2$. In particular, the set of invariant monomials is called the \emph{invariant} sector; all others are called \emph{twisted} sectors.

In \cite{Dijkgraaf2008a} the authors defined two type orbifold partitions. We only review the first type:
\begin{defn}[Orbifold partition of the first type]
It is an ordinary two dimensional partition, with some subset of its boxes distinguished; these distinguished boxes, as points in $\Z^2$ lattice, correspond to monomials with a definite transformation property under the action $(x, y)\rightarrow (\omega x, \bar{\omega}y)$, where $\omega=e^{2\pi i/l}$; we define a weight of such a partition as the number of these distinguished boxes.
\end{defn}
%The orbifold partition is different with the ordinary one. We can see the difference from the Frobenius partition. It is well known that an ordinary two-dimensional partition $\lambda=(\lambda_1, \lambda_2, \cdots, \lambda_k)$ forms a Young diagram $R$. We introduce two sequences of numbers $(a_i)$ and $(b_j)$ by 
%\begin{equation}
%a_i=R_i-i, \ \ \ \ b_j=R_j^t-j,\label{eq2}
%\end{equation}
%where $R^t$ is the dual partition defined by the transposition of the Young diagram $R$. In fact if we slice the Young diagram $R$, $a_i$ and $b_j$ are the number of boxes in rows to the right and to the left of the diagonal respectively:
%\begin{equation}
%\lambda=\left(\begin{array}{cccc}
%a_1 & a_2 & \cdots & a_{d(R)}\\
%b_1 & b_2 & \cdots & b_{d(R)}\end{array}\right),\label{eq3}
%\end{equation}
%where $d(R)$ is the number of boxes on the diagonal. For the ordinary partitions we require that $a_i$ and $b_i$ are strictly decreasing. But for the generalized $\Z_l-$partition we replace it by the following conditions: $a_i$ and $b_i$ are non-increasing and a given number can occur at most k times \cite{Dijkgraaf2008a}. The numbers $a_i$ and $b_i$ are called \emph{generalized Frobenius partitions} \cite{Andrews1984}. 
In \cite{Dijkgraaf2008a} the authors also pointed out that this kind of orbifold partitions are related to states of a Fermi sea. 

We define the generating functions of the generalized partitions of the first type for ALE spaces of $A_{l-1}$ type by 
\begin{equation}
\mathcal{Z}^l_{r, orbifold}=\sum_{\text{first type orbifold partitions}}q^{\#(\text{black boxes})},\label{eq4}
\end{equation}
where $r=0, \cdots, l-1$ and $r$ specifies the power of $\omega$ in the action (\ref{eq1}). 

The orbifold partitions of the first type can be identified with a \emph{blended} partition \cite{Dijkgraaf2008a}. Let us review the definition of the blended partition \cite{Nekrasov:2003rj, Jimbo1983}\footnote{The definition of blended partition can be found in Appendix \ref{A}. The relationship between the blended partition and free fermions can be found in \cite{Dijkgraaf2008a}.}. Consider a colored partition $\vecR=\{R_0, \cdots, R_{l-1}\}$ with charges $p_i, i=0, \cdots, l-1$. The blended partition $K=(K_i)_{i\in\mathbb{N}}$ is defined by the set of integers 
\begin{equation}
\{p+K_m-m|m\in\mathbb{N}\}=\{k(R_{i,m}-m+p_i)+i|i=0, \cdots, l-1, m\in\mathbb{N}\}.\label{eq5}
\end{equation}
The total number of boxes of $K$ is 
\begin{equation}
|K|=\sum_{i=1}^l\left(l|R_i|+\frac{l}{2}p_i^2+ip_i\right)-\frac{(l+1)p}{2}-\frac{p^2}{2},\label{eq6}
\end{equation}
where $p=\sum_{i=0}^{l-1}p_i$. In Appendix \ref{C}, we'll show that $\vecR$ is just the l-core of K, where K is placed at point $(p, 0)$.

In \cite{Dijkgraaf2008a}, after claiming that the orbifold $\Z_l-$partitions of the first type are in one-to-one correspondence with the blended partition $K$ obtained from k-colored partition $\vecR$, such that 
\begin{itemize}
\item {an orbifold $\Z_l-$partition has the same shape as the corresponding blended partition,}
\item{a weight of an orbifold partition (as given by the number of distinguished boxes it contains) is specified by the total weight of a state of k fermions related to $\vecR$,}
\end{itemize}
the authors get the following formula about the number of distinguished boxes 
\begin{equation}
\frac{|K|-\sum_{i=0}^{l-1}ip_i+\frac{(l+1)ln}{2}+\frac{(r+1)r}{2}}{l}=\frac{\sum_{i=0}^{l-1}(l|R_{(i)}|+\frac{l}{2}p_i^2)-\frac{(l+1)p}{2}-\frac{p^2}{2}+\frac{(l+1)ln}{2}+\frac{(r+1)r}{2}}{l}, \label{eq10}
\end{equation}
where the total charge is $p=nl+r, 0\leq r \leq l-1$. Then the partition function (\ref{eq4}) becomes
\begin{eqnarray}
\mathcal{Z}^l_{r, orbifold}&=&\sum_{\{R_{(0)}, \cdots, R_{(l-1)}\}}q^{\sum_i|R_{(i)}|}\sum_{\{p_i|\sum_{i=0}^{l-1} p_i=p=ln+r\}}q^{\frac{\frac{l}{2}p_i^2-\frac{(l+1)p}{2}-\frac{p^2}{2}+\frac{(l+1)ln}{2}+\frac{(r+1)r}{2}}{l}}\nonumber\\
&=&\frac{q^{l/24}}{\eta(q)^l}\sum_{n_1,\cdots, n_{l-1}}q^{\sum_{i}(n^2_i-n_in_{i+1})+\frac{r^2}{2}+n_1r-\frac{r}{2}}\label{eq7}\\ 
&=&\frac{q^{\frac{l}{24}+\frac{r^2}{2l}-\frac{r}{2}}}{\eta(q)}\chi_r^{\widehat{su}(l)_1}(0),\nonumber
\end{eqnarray}
where $\eta(q)=q^{\frac{1}{24}}\prod_{n=1}^\infty(1-q^n)$ is the Dedekind eta function and $n_1, \cdots, n_{l-1}\in \mathbb{N}^{l-1}$ are defined by
\begin{equation*}
(p_0, \cdots, p_{l-1})=\left(\begin{array}{ccc}
n+n_1+r\\
n-n_1+n_1\\
n-n_2+n_3\\
\vdots\\
n-n_{l-2}+n_{l-1}\\
n-n_{l-1}
\end{array}   \right)\in\Z^l.
\end{equation*}

.

\section{Instantons on $A_{l-1}$ ALE space}

In this section we'll review the ADHM construction of instantons on the $\mathbb{C}^2$ and $A_{k-1}$ ALE space \cite{Kronheimer1990, Fucito2004, Fucito2006a, Fujii2007}. % and explain how to get the formula (\ref{eq7}) using the so-called \emph{l-quotient} and \emph{l-core} of Young diagrams. The detailed definitions of l-quotient and l-core of Young diagrams can be found in Appendix \ref{B}. 

\subsection{ADHM on $\C^2$}
The moduli space of U(1) instantons with instanton number k on $\C^2$ has a very beautiful description--the ADHM description. Basically the moduli space is a $U(k)$ quotient of a hypersurface on $\C^{2k^2+2k}$ defined by the ADHM constraints 
\begin{eqnarray}
&&[B_1, B_2]+IJ=0\\
&&[B_1, B_1^\dag]+[B_2, B_2^\dag]+II^\dag-J^\dag J=\xi\mathbf{1}_{k\times k},
\end{eqnarray}
where $B_\alpha, \alpha=1,2$ are linear transformations from a k-dimensional complex vector space $V$ to itself, $I$ is a linear map from $V$ to a 1-dimensional complex vector space $W$ and $J$ is a linear map from $W$ to $V$. So $B_\alpha, \alpha=1,2$ are $k\times k$ matrices, $I$ is a $k\times 1$ matrix and $J$ is a $1\times k$ matrix. The U(k) action is defined by 
\begin{equation}
B_\alpha\mapsto UB_\alpha U^\dag, \ \ \ I\mapsto UI, \ \ \  J\mapsto JU^\dag,
\end{equation}
for $U\in U(k)$. There is a $U(1)^2$ action on $\C^2$. It also induces a $U(1)^2$ on the instanton moduli space. According to \cite{Nakajima:1999}, the fixed points of $U(1)^2$ action on the instanton moduli space is in one-to-one correspondence with the Young diagram. The tangent space of the instanton moduli space is 
\begin{equation}
T\mathcal{M}_k=V^*\otimes V\otimes(Q-\wedge^2Q-1)+W^*\otimes V+V^*\otimes W\otimes \wedge^2Q,
\end{equation}
where $Q$ is a two-dimensional $U(1)^2$ module. 

\subsection{ADHM on $A_{l-1}$ ALE space}
The $A_{l-1}$ type ALE space is defined by the blowup of the quotient $\C^2/\Gamma$ where $\Gamma$ is the $\Z_l$ action:
\begin{equation}
\Gamma: 
\begin{pmatrix}
z_1 \\
z_2
\end{pmatrix}\mapsto 
\begin{pmatrix}
e^{2\pi i/l} & 0\\
0& e^{-2\pi i/l}
\end{pmatrix}
\begin{pmatrix}
z_1\\
z_2
\end{pmatrix}.\label{eq8}
\end{equation}
In fact the U(1) instantons on $\C^2/\Gamma$ can be described by the set $\{(Y, r)\}$ where $Y$ is a Young diagram with $l$ types of boxes and $r$ is an integer $mod\  p$. The integer $r$ specifies the $\Z_l$ representation $R_r$ under which the first box in $Y$ transforms. We know that there exists a $U(1)^2$ action on the instantons on $\C^2$. The specified $r$ gives rise to an embedding of $\Gamma$ into $U(1)^2$:
\begin{equation}
\Gamma: R_a\mapsto e^{2\pi ia/l}R_a, \ \ \ T_1\mapsto e^{2\pi ia/l}T_1, \ \ \ T_2\mapsto e^{-2\pi ia/l}T_2.
\end{equation}
Under this action $V$ and $W$ have the following decomposition: 
\begin{eqnarray}
V&=&\sum_{a=0}^{l-1} V_a\otimes R_a, \ \ \ \text{ dim}V_a=k_a,\\
W&=&\sum_{a=0}^{l-1} W_a\otimes R_a,  \ \ \ \text{ dim}W_a=N_a.
\end{eqnarray}
Since we are dealing with the U(1) instantons, only one of $N_a$ is non-vanishing, and it depends on the $a$ specified. Thus the tangent space of the instanton moduli space is given by the $\Gamma-$invariant component of the $\C^2$ result \cite{Fucito2004, Fucito2006a}
\begin{equation}
T\mathcal{M}_Y=\left(V^*\otimes V\otimes (Q-\wedge^2Q-1)+W^*\otimes V+V^*\otimes W\otimes\wedge^2Q\right)^\Gamma.
\end{equation}
The dimension of the instanton moduli space is
\begin{eqnarray}
\text{dim}_\C\mathcal{M}_Y&=&(k_ak_{a+1}+k_ak_{a-1}-2k_a^2+sk_aN_a)\\
&=&-\hat{C}_{ab}k_ak_b+2k_aN_a,
\end{eqnarray}
where $\hat{C}_{ab}=2\delta_{ab}-\delta_{a,b+1}-\delta_{a,b-1}$ is the extended $A_{l-1}$ Cartan matrix. 

Now let us consider the tautological bundle \cite{Kronheimer1990}. We know the $A_{l-1}$ singularity is resolved by replacing the singularity with $l-1$ intersecting $\mathbb{P}^1$. This leads to new self-dual connections with non-trivial fluxes along the exceptional divisors. The U(1) bundles $\Upsilon^a, a=0, \cdots, l-1$ carry the unit of flux though the exceptional divisors. $\Upsilon^0$ is the trivial bundle. These  $\Upsilon^a, a=0, \cdots, l-1$ have the following property:
\begin{equation}
\int c_1(\Upsilon^a)\wedge c_1(\Upsilon^b)=-C^{ab}, \ \ \ \int_{C_a} c_1(\Upsilon^b)=\delta^b_a,
\end{equation}
where $C^{ab}$ is the inverse of the $A_{l-1}$ Cartan matrix and $C_a$ is the a$^{th}$ exceptional divisor. The gauge bundle $F_Y$ is given by \cite{Kronheimer1990, Fucito2006a}
\begin{equation}
F_Y=\left(V^*\otimes \Upsilon\otimes(Q-\wedge^2Q-1)+W^*\otimes \Upsilon\right)^\Gamma,
\end{equation}
where $\Upsilon=\sum_{a=0}^{l-1}\Upsilon^aR_a$ is the tautological bundle. 

The Chern characters are given by \cite{Fucito2006a}
\begin{eqnarray}
ch_1(F_Y)&=&\sum_a u_ach_1(\Upsilon^a),\\
ch_2(F_Y)&=&\sum_au_ach_2(\Upsilon^a)-\frac{K}{l}\Omega,
\end{eqnarray}
where $K=\sum_ak_a$, $\Omega$ is the normalized volume form of the manifold and 
\begin{equation}
u_a:=N_a+k_{a+1}+k_{a-1}-2k_a=N_a-\hat{C}_{ab}k_b.
\end{equation}
Further, the instanton number $k\in\frac{1}{2k}\Z$ is defined by 
\begin{equation}
k=-\int_M ch_2(F_Y)=\frac{1}{2}\sum_a C^{aa}u_a+\frac{K}{k}=k_0+\frac{1}{2}\sum_a C^{aa}N_a,
\end{equation}
where $C^{aa}=\frac{1}{l}(l-a)a$. 

According to \cite{Vafa1994, Fucito2006a}, the partition function is given by 
\begin{equation}
\mathcal{Z}(q, z_a)=\sum_{k, u_a}\chi(\mathcal{M}_{k, u_a})q^ke^{-z^au_a},\footnote{In fact, we only use the $r$ sector of this partition function. The detailed definitions of the $r$ sector can be found in next section.}
\end{equation}
where $\chi(\mathcal{M}_{k, u_a})$ is the Euler number of the instanton moduli space with first and second Chern characters $u_a$ and $k$ respectively. 

\section{Regular and fractional instantons}
According to \cite{Fucito2004, Fucito2006a, Fujii2007}, the partition function of instantons on $A_{l-1}$-ALE space can be factorized into a product of contributions of regular and fractional instantons. We define the $r$ sector of the partition function $\mathcal{Z}(q, z_a=0)$ by 
\begin{equation}
\mathcal{Z}_r^l=\mathcal{Z}_{reg}\mathcal{Z}_{r, frac}^l,
\end{equation}
where the detailed definitions of $\mathcal{Z}_{reg}$ and $\mathcal{Z}_{r, frac}^l$ can be found in the subsection \ref{sub41} and  subsection \ref{sub42}. 
We'll show that 
\begin{equation}
\mathcal{Z}_r^l\equiv \mathcal{Z}_{r, orbifold}^l.
\end{equation}

\subsection{Regular instantons}\label{sub41}
The regular instantons are instantons in the regular representation of $\Gamma=\Z_l$. They are free to move on $\C^2/\Gamma$. The moduli space is 
\begin{equation}
\mathcal{M}^{reg}_{kl}=(\C^2/\Z_l)^k/S_k,
\end{equation}
which is related to the Hilbert scheme of k-points on $\C^2/\Gamma$ via the Hilbert-Chow morphism. The first Chern class of the regular instanton moduli space is vanishing. Further, according to \cite{Fucito2004, Fucito2006a, Fujii2007}, the regular instantons correspond to the l-quotients $\vecR$ of Young diagrams whose definition can be found in Appendix \ref{B}. Thus the instanton number k has following relationship with the number of boxes of the l-quotients $\vecR=(R_0, R_1, \cdots, R_{l-1})$ of Young diagram $K$ \cite{Fucito2006a}:
\begin{equation}
k=\sum_{i=0}^{l-1}|R_i|.
\end{equation}
Hence it is not hard to find that the partition function of the regular instantons is \cite{Fujii2007}
\begin{equation}
\mathcal{Z}_{reg}=\sum_kq^k \chi(\mathcal{M}^{reg}_{kl})=\frac{q^{\frac{l}{24}}}{\eta(q)^l},%\footnotemark[1]
\footnote{Comparing the q here with notations in theorem 4.6 in \cite{Fujii2007}, our q is equal to $\mathfrak{t}_{0}$ and $\mathfrak{q}=\mathfrak{t}_i=1, \text{ for }i\neq 0\text{ mod }l$. }\label{eq11}
\end{equation}
%\footnotetext[1]
where $\chi(\mathcal{M}^{reg}_{kl})$ is the Euler number of $\mathcal{M}^{reg}_{kl}$. Further, the formula (\ref{eq11}) is consistent with the formula (\ref{eq7}) and the Young diagram $K$ is just the blended partition of $\vecR$. In Appendix \ref{C}, we show that the $\vecR$ is just the l-quotient of the blended partition $K$ of $\vecR$. 

\subsection{Fractional instantons}\label{sub42}
According to \cite{Fucito2004, Fucito2006a, Fujii2007}, the fractional instantons correspond to the Young diagrams which do not have any boxes whose hook length $\ell(s)$ satisfies $\ell(s)=0 \ mod \ l$. In fact, the fractional instantons correspond to the l-cores of Young diagrams whose definition can be found in Appendix \ref{B}. Following \cite{Fujii2007}, we define the partition function $\mathcal{Z}_{r, frac}^{l}$ by
\begin{equation}
\mathcal{Z}_{r, frac}^{l}:=\sum_{Y\in \mathcal{C}^{(l)}} q^{N_{kl-r}},
\end{equation}
where $\mathcal{C}^{(l)}$ is the subset of all Young diagrams consisting of l-cores whose definition can be found in Appendix \ref{B}. According to \cite{Fujii2007}, we have the following formula:
\begin{eqnarray}
\mathcal{Z}_{r, frac}^{l}&=&\sum_{\{p_i|\sum_i p_i=0\}}q^{\frac{1}{2}\sum_{i=0}^{l-1}p_i^2+\sum_{i=l-r}^{l-1}p_i} \\
&=& \sum_{\{\tilde{p}_i|\sum_i \tilde{p}_i=p=ln+r\}}q^{\frac{\frac{l}{2}\sum_i\tilde{p}_i^2-\frac{(l+1)p}{2}-\frac{p^2}{2}+\frac{(l+1)ln}{2}+\frac{(r+1)r}{2}}{l}},
\end{eqnarray}
where the definition of $\tilde{p}_i$ and the detailed derivation can be found in the equations (\ref{eqB2}, \ref{B8}, \ref{B9}) in Appendix \ref{B}.  
According to the Proposition 2.28 in \cite{Fujii2007}, the blended partition $K$ is uniquely determined by the pair (l-quotient, l-core). Hence counting the blended partitions is equivalent to counting the pair of (l-quotient, l-core). Furthermore, according to \cite{Dijkgraaf2008a}, the shape of blended partition is the same as the orbifold partitions. So we have  
\begin{equation}
\mathcal{Z}_{r, orbifold}^l\equiv \mathcal{Z}_r^l=\mathcal{Z}_{reg}\mathcal{Z}_{r, frac}^l.
\end{equation}

\section{Conclusion}
In this note, we have investigated the detailed relationship between the orbifold partitions of first type introduced in \cite{Dijkgraaf2008a} and the pair (l-quotient, l-core) introduced in \cite{Fujii2007}. We find that orbifold partition counting presented in \cite{Dijkgraaf2008a} is exactly the same as counting the (l-quotient, l-core) pair. According to \cite{Fucito2006a, Fujii2007} the U(N) partition function factorizes into 
\begin{equation}
\mathcal{Z}_{U(N)}=\mathcal{Z}^N_{U(1)}=\left(\mathcal{Z}_{U(1), reg}\mathcal{Z}_{U(1), frac}\right)^N.
\end{equation}
It would be interesting to see how to generalize the orbifold partition counting to the U(N) case. 

\acknowledgments{The author wants to thank Prof. Jack Gegenberg for his support of this work. }

\appendix

\section{Colored partitions , charged partitions and blended colored partitions}\label{A}
In this Appendix, we shall review the definitions of colored partitions, charged partitions and blended partitions in \cite{Nekrasov:2003rj}. 

\begin{defn}[Colored partition]
The colored partition $\vecR$ is the l-tuple of partitions: 
\begin{equation}
\vecR=(R_0,\cdots, R_{l-1}),
\end{equation}
where 
\begin{equation}
R_k=(R_{k,1}\geq R_{k,2}\cdots\geq R_{k,n_k}>R_{k, n_{k}+1}=0=\cdots).
\end{equation}
\begin{equation}
|\vecR|:=\sum_{k,i}R_{k,i}
\end{equation}
\end{defn}
\begin{defn}[Charged partition]
The charged partition $(p, \vecR)$ is the set of non-increasing integers $\mathcal{R}_i=R_i+p$, where 
\begin{equation}
\vecR=(R_0\geq R_1\geq\cdots),
\end{equation}
is a partition, and $p\in\Z$. The limit $\mathcal{R}_\infty\equiv p$ is called the charge. 
\begin{defn}[Blending of colored partition]
Given a vector $\vec{p}=(p_0, \cdots, p_{l-1})$, with 
\begin{equation}
\sum_{i=0}^{l-1}p_i=p
\end{equation}
and an N-tuple of partition $\vecR$, we define the \emph{blended} partition K, as follows:
\begin{equation}
\{p+K_m-m|m\in\mathbb{N}\}=\{k(R_{i,m}-m+p_i)+i|i=0, \cdots, l-1, m\in\mathbb{N}\}.
\end{equation}
\end{defn}

\end{defn}

\section{Quotients and cores for Young diagrams}\label{B}
In this appendix, we shall review the definitions of quotients and cores for Young diagrams \cite{Fujii2007, Fucito2006a, Nagao2009}. 

\begin{defn}[Maya diagram]\label{mayadefn}
A Maya diagram is a sequence $\{\mu(k)\}_{k\in\Z}$ which consists of 0 or 1 and satisfies the following property: there exist $N, M\in\Z$ such that for all $k>N$ (resp. $k<N$), $\mu(k)=1$ (resp. $\mu(k)=0$). 
\end{defn}
There is a one-to-one correspondence between the set of Maya diagrams and the set of Young diagrams. We can place the Young diagram Y on the (x, y)-plane by the following way: the bottom-left corner of the Young diagram is at the origin (0,0) and a box in the Young diagram is an unit square. We call an upper-right borderline of $\{x-axis\}\cup\{y-axis\}\cup Y$ the \emph{extended borderline} of Y and denote it by $\partial Y$. The line defined by $y=x$ is called the \emph{medium}. 
\begin{figure}[t]
\centering
\includegraphics[width=0.6\textwidth]{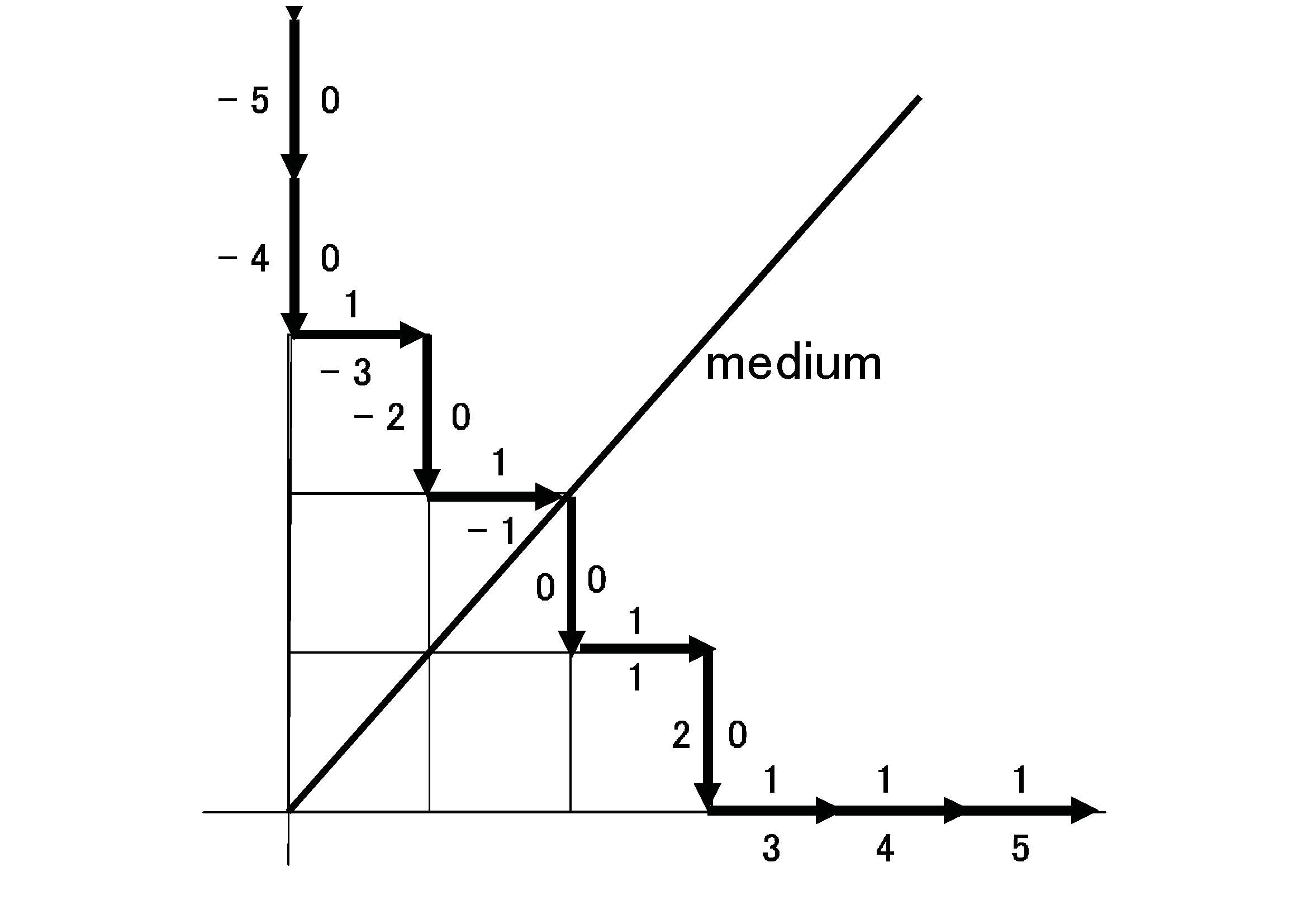}
\caption{Young diagram and Maya diagram copied from \cite{Fujii2007}}
\label{fig:maya}
\end{figure}
The Maya diagram $\{\mu(k)\}_{k\in\Z}$ corresponds to a Young digram Y is defined as follows. We give the direction to the extended borderline $\partial Y$ of $Y$, which goes from $(0, +\infty)$ to $(+\infty, 0)$ (see Figure \ref{fig:maya}). Then each edge of the extended borderline is numbered by $k\in\Z$, if we set the edge which is located at the next to the medium to be $0$. Next we encode a edge $\downarrow$ (resp. $\uparrow$) to 0 (resp. 1). By this way, we have a 0/1 sequence $\{\mu(k)\}_{k\in\Z}$, where each $\mu_Y(k)$ corresponds to a edge of $\partial Y$. 
\begin{defn}[Quotients] \label{lquotient}
For each $0\leq i \leq l-1$, we define 
\begin{equation}
\mu_{Y_i^*}(k):=\mu_Y(lk+i), k\in\Z.
\end{equation}
Then we have an p-tuple $\left(\{\mu_{Y_0^*}(k)_{k\in\Z}\}, \cdots, \{\mu_{Y_{l-1}^*}(k)_{k\in\Z}\}\right)$ of subsequences $\mu_Y(k)_{k\in\Z}$ and each $\{\mu_{Y_i^*}(k)_{k\in\Z}$ is also a Maya diagram. The p-quotient for Y is the l-tuple of Young diagrams $\vec{Y}:=(Y_0^*, \cdots, Y_{l-1}^*)$ corresponding to the Maya diagrams $\left(\{\mu_{Y_0^*}(k)_{k\in\Z}\}, \cdots, \{\mu_{Y_{l-1}^*}(k)_{k\in\Z}\}\right)$.
\end{defn}
\begin{rem}
Here sometime there does not exist a Young diagram corresponding to the Maya diagram $\{\mu_{Y_i^*}(k)_{k\in\Z}\}$, if so, we denote this Young diagram as the empty set $\emptyset$. 
\end{rem}
We define $p_i(Y)\in\Z (i=0, \cdots, l-1)$ by the following condition:
\begin{equation}
\sharp\{\mu_{Y_i^*}(k)=1|n<p_i(Y)\}=\sharp\{\mu_{Y_i^*}(k)=0|n\geq p_i(Y)\}. \label{chargedef}
\end{equation}
\begin{rem}
\begin{equation}
\sum_{i=0}^{l-1}p_i(Y)=0.
\end{equation}
\end{rem}
\begin{rem}
Notice that here we place the vertex of Young diagram to the origin. If we place the vertex at $(p, 0)$, then we have 
\begin{equation}
\tilde{p}_i=\left\{\begin{array}{ccc}
p_0+n,\\
p_1+n,\\
\vdots,\\
p_{l-m-1}+n,\\
p_{l-m}+n+1,\\
p_{l-m+1}+n+1,\\
\vdots,\\
p_{l-1}+n+1,
\end{array}\right. \label{eqB2}
\end{equation}
where $p=nl+m$ and $\tilde{p}_i$ is the new one satisfying (\ref{chargedef}) in the new Young diagram whose vertex is placing at $(p, 0)$. Hence we have 
\begin{equation}
\sum_{i=0}^{l-1}\tilde{p}_i=p.
\end{equation}
\end{rem}
\begin{defn}[l-core]
A Young diagram is called l-core if its l-quotient $\vec{Y}$ is empty. Denote $\mathcal{C}^{(l)}$ as the subset of all Young diagrams consists of l-cores.
\end{defn}
\begin{defn}
Let $Y^{(l)}$ be the Young diagram obtained by removing as many hooks of length $l$ as possible from Y. It is called the \emph{l-core} of Y.
\end{defn}
\begin{prop}
For any $l\geq 2$, a Young diagram Y is uniquely determined by its l-core $Y^{(l)}$ and l-quotient $\vec{Y}$. Then we have
\begin{equation}
|Y|=|Y^{(l)}|+l|\vec{Y}|.
\end{equation}
\end{prop}
\begin{prop}
For $Y\in \mathcal{C}^{(l)}$ which determined by $(p_0, p_1, \cdots, p_{l-1})$ and $\sum_i p_i=0$, then the number of boxes on $y=x-nl+j$ $N_{nl-j}(Y)$ is
\begin{equation}
\sum_{k\in\Z}N_{kl-j}(Y)=\frac{1}{2}\sum_{i=0}^{l-1}p_i(Y)^2+\sum_{i=l-j}^{l-1}p_i(Y). \label{eqB3}
\end{equation} 
\end{prop}
%So when $r=0$ we have
%\begin{equation}
%\sum_{n\in\Z}N_{nl}(Y)=\frac{1}{2}\sum_{i=0}^{l-1}p_i(Y)^2, \label{eqB1}
%\end{equation}
%which is same as the formula (\ref{eq10}), up to $\sum_i|R_{(i)}|$, when $n=r=0$. 

If we place the Young diagram at $(p, 0)$, where $p=ln+r (0\leq r\leq l-1)$, then using (\ref{eqB2}) to replace $p_i$ by $\tilde{p}_i$ in the formula (\ref{eqB3}), then we have the following formula
\begin{eqnarray}
\sum_{k\in\Z}N_{kl-j}(Y)&=&\frac{1}{2}\left\{\sum_{i=0}^{l-j-1}(
\tilde{p}_i-n)^2+\sum_{i=l-j}^{l-1}(\tilde{p}_i-n-1)^2\right\}+\sum_{i=l-j}^{l-1}(\tilde{p}_i-n-1)\nonumber\\
&=&\frac{1}{2}\left\{\sum_{i=0}^{l-j-1}(
\tilde{p}_i^2-2n\tilde{p}_i+n^2)+\sum_{i=l-j}^{l-1}\left[\tilde{p}_i^2-2(n+1)\tilde{p}_i+(n+1)^2\right]\right\}+\sum_{i=l-j}^{l-1}(\tilde{p}_i-n-1)\nonumber\\
&=&\frac{1}{2}\sum_{i=0}^{l-1}\tilde{p}_i^2-n\sum_{i=0}^{l-1}\tilde{p}_i+\frac{n^2l}{2}+\frac{(2n+1)j}{2}-(n+1)j\nonumber\\
&=&\frac{1}{2}\sum_{i=0}^{l-1}\tilde{p}_i^2-n(nl+r)+\frac{n^2l}{2}-\frac{j}{2}\nonumber\\
&=&\frac{1}{2}\sum_{i=0}^{l-1}\tilde{p}_i^2-nr-\frac{n^2l}{2}-\frac{j}{2}.\label{B8}
\end{eqnarray}•
If $j=r$, then we have 
\begin{eqnarray}
\sum_{k\in\Z}N_{kl-r}(Y)&=&\sum_{i=0}^{l-1}\tilde{p}_i^2-nr-\frac{n^2l}{2}-\frac{r}{2}\nonumber\\
&=&\frac{\sum_{i=0}^{l-1}(\frac{l}{2}\tilde{p}_i^2)-\frac{(l+1)p}{2}-\frac{p^2}{2}+\frac{(l+1)ln}{2}+\frac{(r+1)r}{2}}{l}, \label{B9}
\end{eqnarray}•
where $p=ln+r$. 

\section{Blended partition, l-quotient and l-core}
\label{C}
Suppose we have a blended partition K as follows:
\begin{equation}
\{p+K_m-m|m\in\mathbb{N}\}=\{l(R_{i,m}-m+p_i)+i|i=0, \cdots, l-1, m\in\mathbb{N}\}.\label{eqC1}
\end{equation}
%In this section, we'll show that the N-tuple of partition $\vecR=(R_0,\cdots, R_{l-1})$ is just the l-quotient of K, where K is placed at $(p, 0)$.
According to the l-quotient's definition (\ref{lquotient}), it is easy to find that the right hand side of the formula (\ref{eqC1}) tells us the l-quotient of K is $\vecR=(R_0, \cdots, R_1)$ and the positions of 0 are $(R_{i,m}-m+p_i, 0)$ in Maya diagram corresponding to the Young diagram $R_i$. Thus it is not hard to find that $R_i$ is placed at $(p_i, 0)$ position. Further, the left hand side tells us that the positions of 0 of the Maya diagram of K are $(p+K_m-m, 0)$. It implies that the Young diagram K is placed at $(p, 0)$.

Moreover, since the decomposition of K into a pair (l-quotient, l-core) is unique, so it is not hard to imagine that counting the orbifold partition K is equivalent to counting pairs (l-quotient, l-core).

\begin{figure}[t]
\centering
\leavevmode
\includegraphics[width=0.9\textwidth]{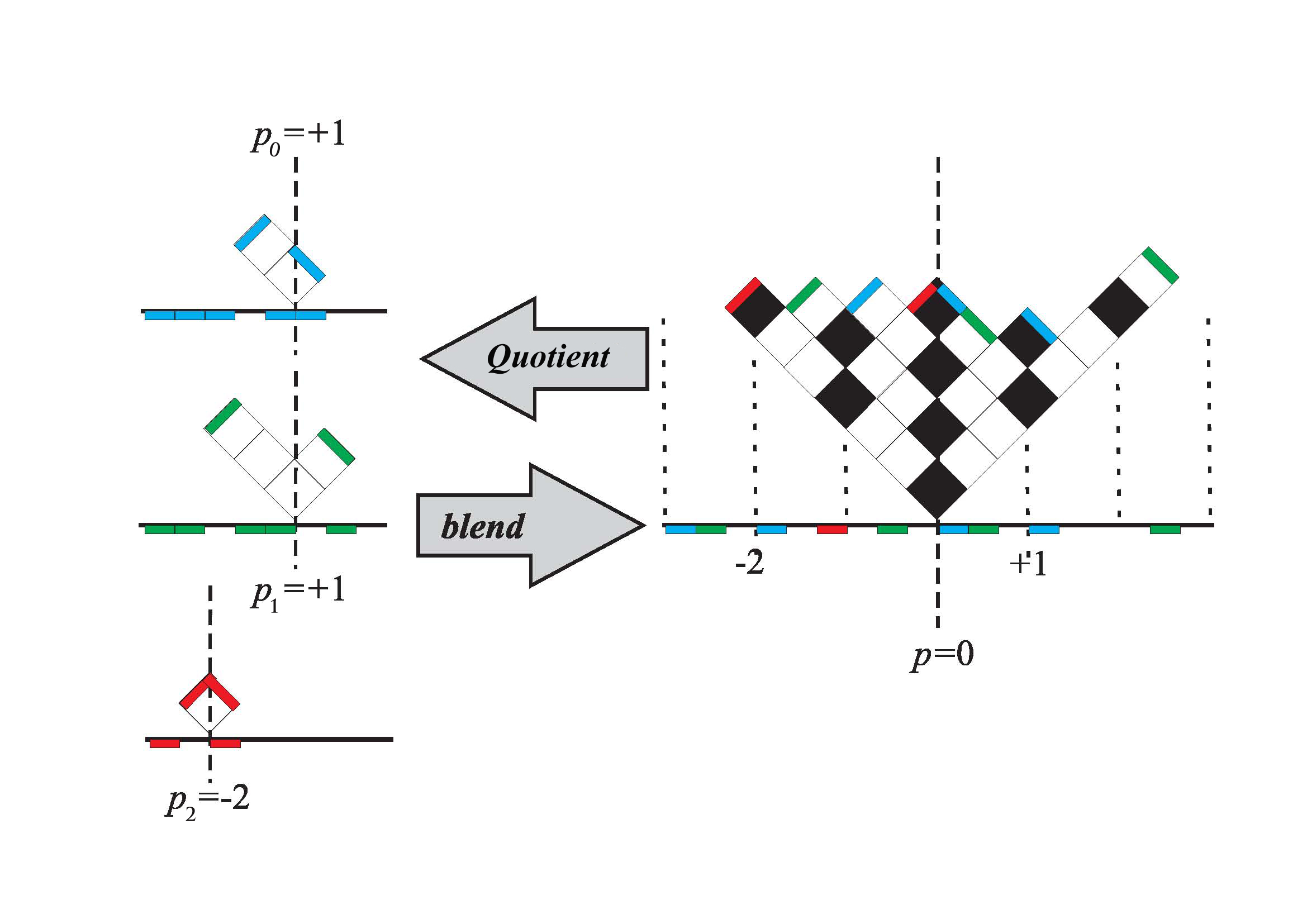}
\caption{The blended partition copied from \cite{Dijkgraaf2008a}.}
\label{fig:blendedpartition}
\end{figure}

Now let us see an example. Figure \ref{fig:blendedpartition} is obtained from \cite{Dijkgraaf2008a}. It shows the blended partition coming from k=3 fermions. Notice the total charge $p=p_0+p_1+p_2=0$.  Table \ref{table1} shows the Maya diagrams corresponding to this blended partition. 
\begin{table}[htd]
\centering
\begin{tabular}{|l|| l l l l l l l l | l l l l l l l l l | }
\hline
 K &\underline{0} & 1 & 0 &1 & 0&1&0&1&0&0&1&0&1&1&1&0&\underline{1}\\
\hline
$R_0$&\underline{0} & & 0& & &1& & &0& & &0& & &1& &\underline{1} \\
\hline
$R_1$&\underline{0}& & & 1& & &0 & & &0 & & &1 & & &0 & \underline{1}\\
\hline
$R_2$&\underline{0}& 1& & & 0& & &1 & & & 1& & &1 & & & \underline{1}\\
\hline
\end{tabular}
\caption{Maya diagrams with vanishing total charge p=0}
\label{table1}
\end{table}
In Table \ref{table1}, the central vertical line is the position of the origin. It is not hard to see that the diagrams $R_0, R_1, R_2$ are exactly same as the left Young diagrams in Figure \ref{fig:blendedpartition}. It also shows the correct position of the vertices of $R_0, R_1, R_2$.  Now let us examine the situation of $p\neq 0$. For simplicity, let us assume $p=1$, so that the Maya diagram is changed to one shown in Table \ref{table2}.
\begin{table}[htd]
\centering
\begin{tabular}{|l|| l l l l l l l |l  l l l l l l l l l | }
\hline
 K &\underline{0} & 1 & 0 &1 & 0&1&0&1&0&0&1&0&1&1&1&0&\underline{1}\\
\hline
$R_0$&\underline{0} & & 0& & &1& & &0& & &0& & &1& &\underline{1} \\
\hline
$R_1$&\underline{0}& & & 1& & &0 & & &0 & & &1 & & &0 & \underline{1}\\
\hline
$R_2$&\underline{0}& 1& & & 0& & &1 & & & 1& & &1 & & & \underline{1}\\
\hline
\end{tabular}
\caption{Maya diagrams with $p=1$}
\label{table2}
\end{table}
It shows that $p_2=-1$ and $p_0, p_1$ does not change; which means the $R_2$ is placed at $(-1, 0)$. Now if we blend $R_0, R_1, R_2$ to a single Young diagram by using the formula (\ref{eqC1}), it is easily to see that this blended Young diagram is exactly same as the $K$ here. 

The 3-core of the Young diagram K in figure \ref{fig:blendedpartition} is as follows:
%{\begin{equation}
%\Yvcentermath1
%\yng(1,1,2,2)
%\end{equation}}
\begin{equation}
\begin{Young}
$b$\cr
\cr
&$b$\cr
$b$&\cr
\end{Young},
\end{equation}
where $b$ stands for black color.

\bibliographystyle{JHEP}
\bibliography{instantons}
\end{document}